\documentclass[multicol,aps,prl,onecolumn,amssymb,showpacs,floatfix]{revtex4}
\usepackage{eucal,bm,graphicx}
\usepackage{gensymb,subfigure}
\usepackage{amsmath}
\usepackage{textcomp}

\begin{document}

  \date{ } \title{Pitfalls in the quantitative analysis of random walks and the mapping of single-molecule dynamics at the cellular scale.} \author{Jean-Baptiste Masson$^{1,2}$, Maxime Dahan$^{3}$ and Antoine Triller$^{4}$.\\ 
  \small{$^{1}$ Physics of Biological Systems, Institut Pasteur, 75724 Paris Cedex 15, France} \\
  \small{$^{2}$Centre National de la Recherche Scientifique, Unite Mixte
de Recherche 3525, 75015 Paris, France} \\
  \small{$^{3}$ Laboratoire Physico-Chimie Curie, Institut Curie, CNRS UMR168, Universite Pierre et Marie Curie-Paris 6, 75005 Paris, France}\\
  \small{$^{4}$ Biologie Cellulaire de la Synapse, Institut Nationale de la Sante et de la Recherche Medicale U789, Ecole normale superieure,
75005 Paris, France.} \\}


\begin{abstract}
In recent years Bayesian Inference has become an efficient tool to analyse single molecule trajectories. Recently, high density single molecule tagging, Langevin Equation modelling and Bayesian Inference \cite{JBM_carte} have been used to infer diffusion, force and potential fields at the full cell scale. In this short comment, we point out pitfalls \cite{holcman_arxiv,Hoze}  to avoid in single molecule analysis in order to get unbiased results and reliable fields at various scales.

\end{abstract}

\maketitle

\textbf{A response to Holcman \textit{et al} \cite{holcman_arxiv}}
\newline
\newline
Understanding the physical and biological processes governing the trafficking of membrane proteins is an central question in cell biology and biophysics. A powerful approach to model the motion of a random walker in a complex environment is to use an overdamped Langevin equation. In the case of a coarse-grained description (corresponding to the limited spatial and temporal resolution accessible in experiments), this equation writes: 
$$
\dot{\bf{r}} = \bf{F}\left(\bf{r}\right)/\gamma\left(\bf{r}\right) + \sqrt{2D\left(\bf{r}\right)}\bf{\eta}\left(t\right)
$$
where $\bf{F}\left(\bf{r}\right)/\gamma\left(\bf{r}\right)$ corresponds to a local drift, $\gamma\left(\bf{r}\right)$ is a spatially-varying friction, $D\left(\bf{r}\right)$ a spatially-varying diffusion coefficient and $\bf{\eta}\left(t\right)$ is a zero-averaged gaussian noise. 
\newline

In past years, the advent of single molecule techniques has opened new possibilities to estimate the values of the local parameters $\bf{F}\left(\bf{r}\right), \gamma\left(\bf{r}\right)$ and $D\left(\bf{r}\right)$ in the cellular context. In particular, bayesian inference tools have been developed to extract local force and diffusivity fields based on single molecule trajectories \cite{JBM1,JBM2,JBM3,JBM4, Voisinne, Richly}. Not only these tools have been extensively characterized from a methodological point of view \cite{JBM1,JBM2,Voisinne} but they have also been applied to experimental systems, in order to investigate the dynamics of toxin receptors in lipid rafts \cite{JBM1,JBM3,JBM4} and to calibrate optical tweezers \cite{Richly}. 
\newline

More recently, the analysis of the force and diffusivity fields has been extended to the full cellular scale in \cite{Hoze} and in \cite{JBM_carte} thanks to high-density single molecule techniques, such as sptPALM \cite{Manley} and uPAINT \cite{Gianonne}. Thereby, it has been possible to quantitatively analyze the synaptic stabilization of AMPA receptors \cite{Hoze} and Glycine receptors \cite{JBM_carte}. Both papers similarly interpreted the role of synaptic scaffolds in terms of local confining potentials in line with a view of the neuronal membrane where neuroreceptors diffuse and transiently stabilize at synaptic sites \cite{Triller}. 
\newline

In their comment, Holcman \textit{et al.} question the validity of the approach and description presented in \cite{JBM_carte}. We reply below to the points that they raised by:  (i) addressing some computational issues (statistical estimation of the model parameters, numerical simulations) and demonstrating biases in the statistical estimators used in \cite{Hoze}, (ii) discussing the interpretation of the results.

\vspace{0.5cm}

\noindent \textbf{Parameter estimation and numerical simulations.}
\newline

A key point for the determination of the force and diffusivity fields based on single molecule data is the choice of the statistical estimators. To clearly illustrate the simplifying assumptions used in \cite{Hoze}, we first remind the standard definition of the simplest likelihood used to analyze the dynamics of individual biomolecules in the absence of positional noise:
$$
L\left(D,\bf{f}\right) = \frac{1}{\left(4\pi D \Delta t \right)^{N}}\exp\left(-\sum^{N}_{i=1}\frac{\left(\Delta \bf{r}_{i} - \bf{f} \Delta t\right)^{2}}{4 D \Delta t}\right)
$$
corresponding to $N$ translocations $\Delta \bf{r}_{i}$ (during the sampling time $\Delta t$) in a subdomain characterised by a constant drift $\bf{f}= \bf{F}/\gamma$ and diffusion coefficient $D$.  By differentiating $\log(L)$ with respect to $\bf{f}$ and $D$, one directly gets:
$$
\frac{\partial \log(L)}{\partial\bf{f}}=\sum^{N}_{i=1}\frac{\left(\Delta \bf{r}_{i} - \bf{f} \Delta t\right)}{2 D}
$$
and
$$
\frac{\partial \log(L)}{\partial D}= - \frac{N}{D} + \sum^{N}_{i=1}\frac{\left(\Delta \bf{r}_{i} - \bf{f} \Delta t\right)^{2}}{4 D^2 \Delta t}
$$
The maxima are obtained for:
$$\bf{f}=(\sum^{N}_{i=1}\Delta \bf{r_{i}})/N\Delta t$$ 
and 
$$ D = \frac{1}{4N\Delta t} \sum_i(\Delta \bf{r}_{i} - \frac{1}{N} \sum_j \Delta \bf{r}_{j})^2 $$
The estimators used in \cite{Hoze} correspond to the maxima in the limit $\bf{f} \Delta t \ll \Delta \bf{r}_{i}$ and can thus be designated as Maximum Likelihood Estimators (MLE).
 \newline

The estimators used in \cite{Hoze} are biased when applied to experimentally measured trajectories for two main reasons. First, an important element in the analysis of single molecule trajectories is the positioning noise $\sigma$, which results from a combination of processes such as photon noise, motion blur, electronic noise and limited performance of peak fitting algorithm. However, the effect of positioning noise was \textbf{entirely neglected} in the analysis  presented in \cite{Hoze}, which, as shown below, leads to estimator biases and to improper parameter values for the synapses studied in \cite{Hoze}. Second, the approximation $\bf{f} \Delta t \ll {\Delta \bf{r}_{i}}$ (not discussed in \cite{Hoze}) also leads to strong bias in the force and diffusion estimation over range of values relevant for the experiments reported in  \cite{Hoze}. In our work, we have used Bayesian inference schemes and directly included in our computation the full positioning noise as an experimental parameter $\sigma$ (see eq. equations 5 and 6 in \cite{JBM_carte}). We refer to \cite{Voisinne} for a demonstration of the optimality of Bayesian inference methods for diffusion measurements and to \cite{JBM3} for a detailed discussion on their use in the case of confined motions.
\newline

To directly compare the performances of the estimators used in \cite{Hoze} and \cite{JBM_carte}, we have performed numerical simulations corresponding to experimental conditions reported in \cite{Hoze} ($\sigma = 25$\,nm, $\Delta t = 20$\,ms, about  40 points per subdomain). We first estimated the diffusion coefficient in the absence of external drift for $D$ varying between 0.01 and 0.5 $\mu$m$^2$/s (Figure 1A). For $D < 0.05 \mu$m$^2$/s (a value comparable to the diffusion coefficient in synaptic domains), we observe a clear bias when using the MLE estimators whereas our estimator remains accurate over the entire range. The limit corresponds to the regime where $\sigma$ is comparable or greater than $\sqrt{D\Delta t}$. 

Next, we tested the estimation of the diffusion coefficient and drift when brownian particles (with a diffusion coefficient $D = 0.05 \mu$m$^2$/s)  were submitted to a force varying between $10^{-2}$ and $10^{1}$ pN. While our approach yielded accurate estimates of $F$ and $D$, there was a systematic bias in the MLE estimation of both the force and the diffusion coefficient at low force. Furthermore, a strong deviation can be observed for forces larger than $F_c \sim 0.1$ pN (Figure 1.B1,B2), a value corresponding to the the limit where the displacement due to the drift $f \Delta t$ is comparable to the diffusional movement $\sqrt{D\Delta t}$. Importantly, $F_c$ is comparable to the confinement force $U_0/2\delta$ in a potential well with depth $U_0$ = 8 k$_B$T and extension $\delta$ = 200 nm, such as the ones reported in \cite{Hoze}. Overall, our simulations indicate that the estimates of the parameters (diffusion, forces and potential) for the synaptic domains are not accurate in \cite{Hoze}

Note that Bayesian inference techniques have been recently applied to evaluate the experimental trapping force in optical tweezers \cite{Richly}. This system is advantageous since the trap parameters can be finely tuned with the light intensity and, importantly, independently calibrated. The estimate of the trap spring constant determined with inference schemes was in good agreement with (in fact more accurate than) the results obtained by standard tools (equipartition and power-spectrum methods). 
\newline

Another important point is the extraction of the potential from the force field. In \cite{Hoze}, this is done by minimizing the square difference between the gradient of the potential and the force field. Such class of optimization problems, especially with a discrete set, limited number of points and harmonic approximation, are \textbf{always solved with a regularization term} \cite{machine_learning,machine_learning_2} (see equation 7 in \cite{JBM_carte}). Solving these problems without any regularization, either of the force field or of the potential, leads to biases (as illustrated in Figure 1C,D). 
\newline

Finally, regarding the numerical simulations (including the Gillespie scheme), we refer to \cite{FP} for approximating the Fokker-Planck by the master equation and to \cite{Gillespie} for simulating the master equation. We especially emphasize the importance of the Gillespie scheme for rapidly generating a large number of trajectories, which enabled us to compute fundamental descriptors of the motion (such as the propagator) as function of the time and over a large range of distance. 

\vspace{0.5cm}

\noindent \textbf{Interpretation and discussion of the model.}
\newline

A legitimate topic of discussion is the interpretation of the diffusivity and force (or potential) fields estimated from individual trajectories.  In their experiments, Hoze \textit{et al} noted the existence of domains where individual translocations pointed toward a single center point \cite{Hoze}. Similarly to what has been reported earlier for toxin receptors \cite{JBM1,JBM3}, such behavior was described in terms of a trapping potential. For AMPA receptors, the potential wells were interpreted as resulting from interactions of the receptors with scaffolding proteins PSD-95 via stargazin molecules.
\newline

Before discussing and comparing our results \cite{JBM_carte},  we emphasize that our inference approach does not require the computation of a diffusion and a potential $(D,U)$ but can be performed to estimate the parameters $(D,F)$ or $(D, F/\gamma)$ as it was done in \cite{JBM1, JBM2, JBM3, JBM4, Richly, Voisinne, silvan}. The inference scheme simply provides a description of the environment in terms of the diffusivity and force fields that have the highest probability. In the specific case of our experiments on glycine receptors \cite{JBM_carte}, we have favored a description in terms of a potential for several reasons. First, the force field did not exhibit rotational terms, thus justifying the computation of a potential. Second,  glycine receptors are known to interact with gephyrin scaffolding proteins at inhibitory synapses. The fact that the trapping areas of the membrane proteins perfectly co-localized with the location of gephyrin clusters strongly supported the notion that the clusters acted as trapping wells. The role of the scaffolds in the observed potentials was further demonstrated experiments using membrane proteins with mutant constructs of the $\beta$-loop mediating the receptor-gephyrin interactions: for a mutant with weaker affinity, we measured shallower traps, and for proteins lacking the $\beta$-loop entirely, there was a complete absence of potentials. In this context, the potential depth represents the energy needed for receptors to escape gephyrin clusters. It is worth noting that it is the classical definition of the potential in physics.
\newline

In \cite{JBM_carte}, we have used a spatially varying friction $\gamma\left(\bf{r}\right)$, reflecting the heterogeneity of the plasma membrane at the full cell scale. Furthermore, we imposed the relationship $D\left(\bf{r}\right) = k_{B}T/\gamma\left(\bf{r}\right)$. This hypothesis was supported by the fact that the transitions between adjacent domains (determined experimentally) satisfied detailed balanced condition, thus suggesting a local equilibrium. For more advanced discussions on the FD relationship for systems with spatially varying friction, we refer to recent references \cite{FD_1,FD_2}. 
\newline

An important difference between the results of \cite{Hoze} and \cite{JBM_carte} lies in the quantitative measurements of the potential depths. For the wild-type beta-loop, we found an average energy of 3.4 k$_B$T with less than 15$\%$ of the traps above 6 k$_B$T (measured over $\sim$ 69 clusters).  In contrast, 25 $\%$ of the traps reported in \cite{Hoze} were deeper than 8 k$_B$T, indicating a much higher level of stabilization. While it could  be due to the difference between the biological systems and their molecular constituents, it seems likely that the biased estimators (see discussion above) and the limited number of data points presented in \cite{Hoze} (2, 5 and 10 wells depending on the experimental conditions) also contribute to the discrepancy. Additional experiments, combined with careful statistical analysis, are needed to clarify the difference between excitatory and inhibitory scaffolds.
 \newline
 
In our view, the main remaining challenge is to quantitatively determine how the trapping energies at scaffolding clusters relate to the biochemical properties of their individual constituents. We emphasize that there is no general procedure to go from a diffusion and force fields to biochemical characteristics of complex molecular assemblies.  It will depend on many geometrical and molecular properties of the assemblies, as well as on the spatio-temporal resolution of the measurements. Fortunately, nanoscopy imaging methods now provide powerful tools to analyze the composition and structure of scaffolding complexes \cite{Specht,Hosy} and relate them to their functional roles at synapses.

\vspace{0.5cm}

\noindent \textbf{Conclusion.} 
\newline

In summary, we stand by the methods and the approach described in \cite{JBM_carte} and earlier papers, and point out to errors in the statistical and computation methods used in \cite{Hoze,holcman_arxiv}. Note also that we did not comment here on other important elements for the mapping of the diffusivity and force (or potential fields), such as the meshing procedure and the mesh size which should be adjusted to the averaged translocation length (a point not discussed in \cite{Hoze}), the effect of strong variation between the number of points between neighboring mesh domains, the effect of confinement on the accuracy of parameters extraction, the effect of heterogeneous diffusion field on the convergence of the estimators and the projection of the potential on a proper functional basis. We refer the reader to previous publications \cite{JBM1, JBM2, JBM3, JBM4, Voisinne, Richly, JBM_carte, silvan} (including their supplementary materials) for further discussion on all these points.

\begin{figure}[tbp]
\begin{center}
\includegraphics[width=500pt]{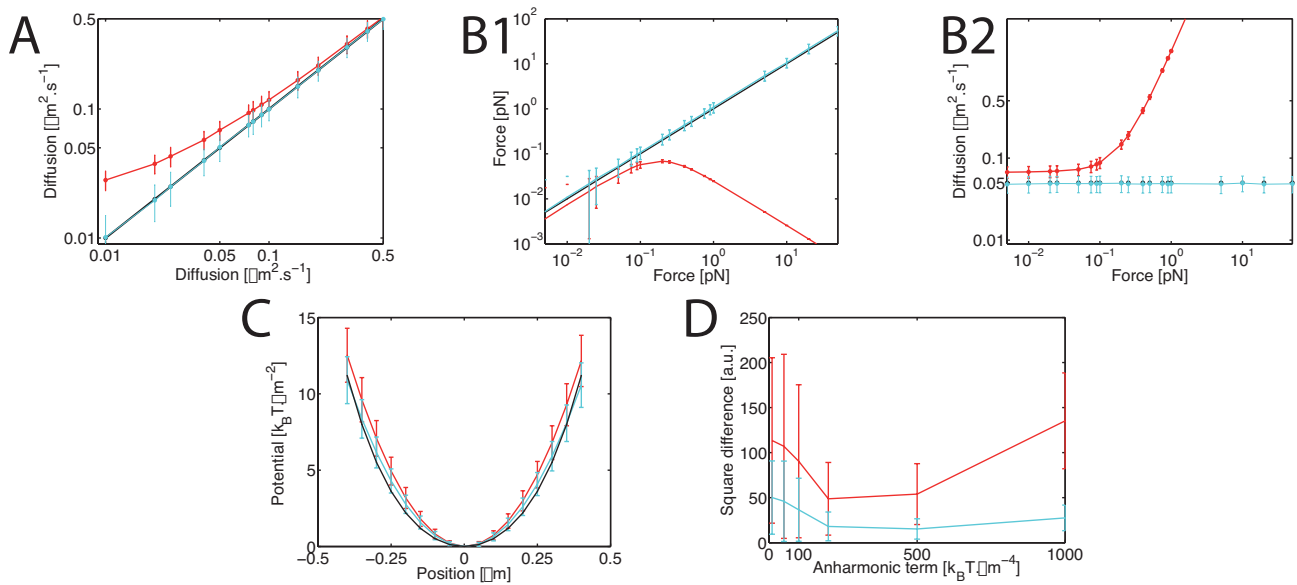}
\end{center}
\vskip -3.8mm
\caption{Comparison of the results of the estimators used in \cite{Hoze} and \cite{JBM_carte} in simulated trajectories. All values are averaged over 1000 trajectories. The theoretical values are in black, the results associated to the estimators described in \cite{Hoze} are in red, and to our estimators in cyan.  Positioning noise was set to 25 nm \cite{Hoze}. \textbf{A} Evolution of the estimated diffusion with constant diffusion field matching experimental conditions in \cite{Hoze}. Note that blue and red error bars are the same size, distortion is due to log-log plot.  \textbf{B1,B2} evolution of the force and diffusivity (respectively) with force for constant diffusion field $D= 0.05\mu m^{2}.s^{-1}$. \textbf{C} Example of an extracted potential from a simple 1D force field. The mesh was 50nm \cite{Hoze}. \textbf{D} evolution of the difference between the input potential and the reconstructed one for 2D force fields, as a function of the anharmonicity.}
\vskip -1.8mm
\label{estimators}
\end{figure}{\em }

\bibliographystyle{apsrev}
\bibliography{masson}
$^{*}$ Corresponding authors: jbmasson@pasteur.fr, maxime.dahan@curie.fr, triller@biologie.ens.fr.

\end{document}